\documentclass[useAMS,usenatbib]{mn2e}
\usepackage{graphics}
\usepackage{graphicx}
\usepackage{psfig}
\input{epsf}

\def\etal{et al.}

\def \lleq {\lower0.9ex\hbox{ $\buildrel < \over \sim$} ~}
\def \ggeq {\lower0.9ex\hbox{ $\buildrel > \over \sim$} ~}

\def\spose#1{\hbox to 0pt{#1\hss}}
\def\simle{\mathrel{\spose{\lower 3pt\hbox{$\mathchar"218$}}
     \raise 2.0pt\hbox{$\mathchar"13C$}}}
\def\simge{\mathrel{\spose{\lower 3pt\hbox{$\mathchar"218$}}
     \raise 2.0pt\hbox{$\mathchar"13E$}}}

\def\beq{\begin{equation}}
\def\eeq{\end{equation}}
\def\ber{\begin{eqnarray}}
\def\eer{\end{eqnarray}}

\def\be{\begin{equation}}
\def\ee{\end{equation}}

\begin{document}

\title[Expansion history of the universe and the properties of dark energy]{Model independent reconstruction of the expansion history of the universe and the properties of dark energy}

\author[Arman Shafieloo]{Arman Shafieloo\thanks{arman@iucaa.ernet.in} \\
%
Inter--University Centre for Astronomy and Astrophysics, Post Bag 4, Ganeshkhind, Pune 411007, India\\
%
  } \maketitle

\begin{abstract}
We have improved upon the method of smoothing supernovae data to reconstruct the expansion history of the universe, $h(z)$, using two latest datasets, Gold and SNLS. The reconstruction process does not employ any parameterisation and is independent of any dark energy model. 
 The reconstructed $h(z)$ is used to derive the distance factor $A$ up to redshift $0.35$ and the results are compared with the given value of $A$ from detection of baryon acoustic oscillation peak (BAO). We find very good agreement between supernovae observations and the results from BAO for $\Omega_{0m} \approx 0.276 \pm 0.023$. The estimated values of $\Omega_{0m}$ are completely model-independent and are only based on observational data. The derived values of $\Omega_{0m}$ are then used to reconstruct the equation of state of dark energy, $w(z)$. Using our smoothing method we can demonstrate that while SNLS data are in very good agreement with $\Lambda$CDM, the Gold sample slightly prefers evolving dark energy. We also show that proper estimation of the equation of state of dark energy at the high redshifts would be impossible at the current status of observations.    
\end{abstract}


\section{Introduction}
Over the last two decades, cosmology has entered the stage of `precision' science, which involve significant improvements in observational techniques,
implementation of more powerful statistical and mathematical tools, and, of
course, greatly advanced computational facilities. While these advancements
have yielded much new insight into the subject, many important questions still
remain unanswered. The nature of dark energy has been the subject of much debate over the past decade \citep{ss00,carrol01,pr03,pad03,s04,cst06}. Supernovae data, which gave the first indication of the accelerated expansion of the universe, are expected to elucidate this interesting question further, as the quality of the data steadily improves \citep{riess98,perl99,knop03,tonry03,riess04,snls05,riess06}.
Much attention in recent years has focused on determining the properties of dark energy in a model independent manner. This can be done using either parametric \citep{star98,huter99,saini00,chiba00,asss03,chev01,weller02,efstathiou,maor02,copeland,linder,wangm,saini04,leandros,tirth05,gong,linderhuterer,huter06,ass07,gw07,barger,ohta1,ohta2,sahlen} or non-parametric methods \citep{hut03,wang01,saini03,dal03,wanteg04,wanteg05,huterer05,sass06,durrer,reza06}. A comprehensive recent review has been given by \citet{ss06}.
An earlier work by \citet{sass06} suggested a non-parametric method
based on smoothing the supernova data over redshift in order to
reconstruct cosmological quantities, including the expansion rate,
$h(z)$, and the equation of state of dark energy, $w(z)$, in a {\it
model-independent} manner. In this approach, the data are dealt with
directly, and one does not rely on a parametric functional form for
fitting any of the quantities $d_{L}(z)$, $h(z)$ or $w(z)$. The result
obtained by using this approach is, therefore, expected to be
model-independent. This method was shown to be successful in
discriminating between different models of dark energy if the quality
of data is commensurate with that expected from the future SuperNova
Acceleration Probe (SNAP). In this paper we improve the smoothing
method and apply it to two recent sets of supernovae data: Gold
\citep{riess06} and SNLS \citep{snls05}. We then compare the derived
expansion history of universe with the results of baryon acoustic peak
observations \citep{bao05}. Specifically, we use the improved
smoothing method to reconstruct the Hubble parameter, $h(z)$, and then
derive the distance factor, $A$, up to a redshift of $0.35$
independently of the assumption of any cosmological model. This
derived value, based on supernovae data, is then compared with the
distance factor $A$ (which is also claimed to be relatively
independent of dark energy model) being determined by the detection of
the baryon acoustic oscillation peak. One of the main results of this
paper is that there is a good agreement between supernovae data (both
Gold and SNLS) and baryon acoustic peak observations for the values
$\Omega_{0m} \approx 0.276 \pm 0.023$. The derived value of
$\Omega_{0m}$ is then used to reconstruct the equation of state of
dark energy for both supernovae datasets. 
We should emphasize here that all the results in this paper are
only based on observational data and no theoretical model has been
assumed. This is an advantage of this method over the functional
fitting methods in which the reconstructed results are biased by an
assumed functional form or a theoretical model. The paper is organised
as follows. In Section II we briefly explain the smoothing method and
we estimate the accuracy of the method based on the quality and the
quantity of current datasets. In Section III we apply the smoothing
method on the Gold supernovae dataset and by using the results of
detection of baryon acoustic oscillation peak, we reconstruct
$w(z)$. In Section IV, we carry out a similar treatment on the SNLS
dataset. Finally, in Section V we discuss our results along with some 
concluding remarks.

\section{Method of smoothing}

The method of smoothing belongs to the category of non-parametric methods of reconstruction which is complementary to the approach of fitting a parametric ansatz to the dark energy density or the equation of state. Most papers using the non-parametric approach exploit a kind of top-hat smoothing in redshift space. Instead, we follow a procedure which is well known and frequently used in the analysis of large-scale structure \citep{coles,martinez}; namely, we attempt to smooth noisy data directly using a Gaussian smoothing function. 
In this method we apply Gaussian smoothing to supernova data (which are of the form $\lbrace \ln d_L(z_i),z_i \rbrace$) in order to extract information about important cosmological parameters such as $H(z)$ and $w(z)$. The smoothing algorithm calculates the luminosity distance at any arbitrary redshift $z$ to be

\ber
\label{eq:bg}
\ln d_L(z,\Delta)^{\rm s}=\ln
\ d_L(z)^g+N(z) \sum_i \left [ \ln d_L(z_i)- \ln
\ d_L(z_i)^g \right] &&\nonumber\\
{\large \times} \ {\rm exp} \left [- \frac{\ln^2 \left
( \frac{1+z_i}{1+z} \right ) }{2 \Delta^2} \right ], &&\\
N(z)^{-1}=\sum_i {\rm exp} \left
[- \frac{\ln^2 \left ( \frac{1+z_i}{1+z} \right ) }{2 \Delta^2}
\right ]~. \hspace{2.8cm}&&\nonumber
\eer

Here $\ln d_L(z,\Delta)^{\rm s}$ is the smoothed luminosity distance
at any redshift $z$ which depends on luminosity distances of each SNe
event with the redshift $z_i$, and $N(z)$ is a normalisation
parameter. The quantity $\ln d_L(z)^g$ represents a guess background model which we subtract from the data before smoothing.  This approach allows us
to smooth noise only, and not the luminosity distance. After noise
smoothing, we add back the guess model to recover the luminosity
distance.  This procedure is helpful in reducing noise in the results.
Since we do not know which background model to subtract, we may take as a
reasonable guess that the data should be close to $\Lambda$CDM and use
$d_L(z)^g=d_L(z)^{\Lambda {\rm CDM}}$ as a first approximation and then use a
boot-strapping method to find successively better guess models.
Having obtained the smoothed luminosity
distance, we differentiate it once to obtain the Hubble parameter,
\be
\label{eq:hubb}
H(z)= \left[ \frac{d}{dz} \left( \frac{d_L(z)}{1+z} \right) \right]^{-1}\,\,,
\ee
and once again to obtain the equation of state of dark energy $w(z)$,
\be
\label{eq:w}
w(z)=\frac{[2 (1+z) /3] \ H^{\prime}/H - 1}{1 \ - \ (H_0/H)^2 \Omega_{0m} \ (1+z)^3}.
\ee

In any kind of smoothing scheme for the luminosity distance, some bias is introduced both in $d_L$ and in derived quantities like $H(z)$ and $w(z)$ (see appendix A1 in \citet{sass06} to find detailed calculations of the bias). 
It is important to choose a value of $\Delta$ which gives a small
value of the bias and also reasonably small errors on derived
cosmological parameters. To estimate the value of $\Delta$ in
(\ref{eq:bg}), we consider the following relation between the
reconstructed results, quality and quantity of the data and the
smoothing parameters. One can show that the
relative error bars on $H(z)$ scale as \citep{tegmark}
\be
\label{eq:tegmark}
\frac{\delta H}{H}\propto \frac{\sigma}{N^{1/2} \Delta^{3/2}}\,\,,
\ee 
where N is the total number of supernovae (for approximately uniform
distribution of supernovae over the redshift range) and $\sigma$ is
the noise of the data. From the above equation we see that a larger
number of supernovae or larger width of smoothing, $\Delta$, will
decrease the error bars on the reconstructed $H$, but as it has been reported earlier \citep{sass06}, the bias of the method is approximately
related to $\Delta^2$. This implies that by increasing $\Delta$ we
will also increase the bias of the results. If we attempt to estimate
$\Delta$ such that $\frac{\delta H}{H}\propto 3\sigma $, then for $N=182$ data points (which is the number of data points in the Gold sample), we get $\Delta=0.084$ for a single iteration of our method. However, with each iteration, the errors on the parameters will increase. Therefore, using this value of $\Delta$ when we use an iterative process to find the guess model will result in such large errors on the cosmological parameters as to render the reconstruction exercise meaningless. It has been shown in \citet{sass06} that at the M-th iteration, the error on $\ln d_L$ will be approximately $\delta_M (\ln d_L) \simeq \sqrt{M} \delta_0 (\ln d_L)$, and the error on $\ln d_L$ scales as $1/\Delta$. We would like the errors after $M$ iterations to be commensurate with the optimum errors obtained for a single iteration, $\delta_{0}$, so we require $\Delta_{\rm optimal} \simeq \sqrt{M} \Delta_{0}$. Therefore, if we wish to stop the boot-strapping
after $50$ iterations, then $\Delta_{\rm optimal} \simeq 0.6$. However, after this rough estimation of the values of $\Delta$ and $M$, we can still play around these values to find the best combination, by minimising the likelihood of the reconstructed results to the data. In the following, we use $\Delta=0.6$ and we calculate the $\chi^2$ of the reconstructed distance moduli to the data after each iteration, and we stop the boot-strapping process after reaching the minimum value of $\chi^2$. This effect, that $\chi^2$ of the reconstructed results goes to a minimum value and increases again with iteration is a reflection of the problem of some iterative reconstruction algorithms which are not error-sensitive. In these cases the noise will be added to the reconstructed results after certain number of iterations and the iterative process should be stopped after reaching the minimum value of $\chi^2$ to get the best result. Similar effect has been reported and studied in the Richardson-Lucy deconvolution algorithm to reconstruct the form of primordial power spectrum from CMB data \citep{arun04}.

In appendix A we show that the results are not sensitive to the chosen value of $\Delta$ and also to the assumed initial guess model.

\section{Results from the Gold Dataset}
The recently released Gold sample \citep{riess06}, consist of 182 supernovae type Ia which have been gathered from five different subsets of data, observed during the last 16 years. The range of redshift for these supernovae are between $0.024$ and $1.75$. In this section we use this dataset to reconstruct $h(z)$, estimate the value of $\Omega_{0m}$, and then reconstruct $w(z)$.
We choose a flat $\Lambda$CDM model with $\Omega_{0m} =0.30$ as the initial guess model in our calculation and we fix the value of $\Delta$ (width of smoothing) to be $0.6$. After each iteration, we compute the $\chi^2$ and we stop the boot-strapping process once $\chi^2$ reaches its minimum value.

The $\chi^2$ at any iteration is calculated from the formula, 
\be
\label{eq:chis}
\chi^2_{rec,j}(H_0) = \Sigma_i \frac{(\mu_{rec,j}(H_0,z_i) - \mu_{obs}(z_i))^2}{\sigma^2_{i}}
\ee 
and is followed by marginalising over $H_0$. We have marginalised over $H_0$ by integrating over the probability density $p \propto exp(-\chi^2 /2)$ for all values of $H_0$.

\begin{figure}
\includegraphics[scale=0.34, angle=-90]{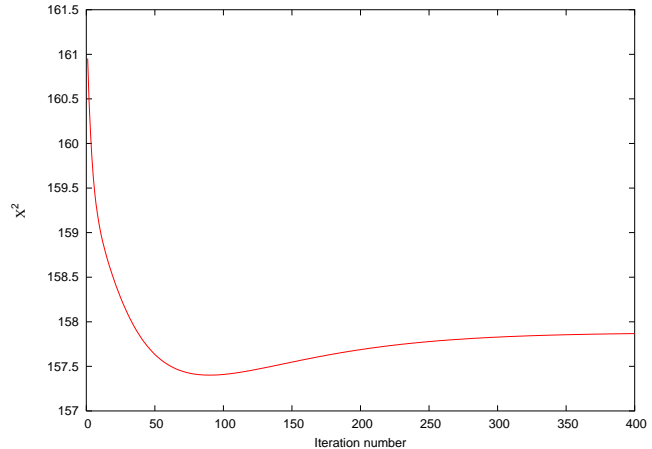}
 \caption{ Computed $\chi^2$ for the reconstructed results at each iteration, using Gold sample.}
\label{chi_gold}
\end{figure}

In Eq.\ref{eq:chis}, $\mu_{rec,j}(H_0,z_i)$ is the reconstructed result at the $j$th iteration for the distance moduli at redshift $z_i$, assuming the value of $H_0$, and $\mu_{obs}(z_i)$ is the Gold sample data given by \citet{riess06}. In Figure \ref{chi_gold} we show the $\chi^2$ of the reconstructed results at different iterations, after marginalising over $H_0$. As we see, the $\chi^2$ has a minimum around $j=89$ and after this, $\chi^2$ is slowly increasing. So we stop the boot-strapping process at this iteration and determine $h(z)$. We can also see that for the initial guess $\Lambda$CDM model, the $\Delta \chi^2$ of the best recovered result is less than $4$ which means that the flat $\Lambda$CDM model is in agreement with the Gold sample within $2\sigma$.    
 
\begin{figure}
\includegraphics[scale=0.34, angle=-90]{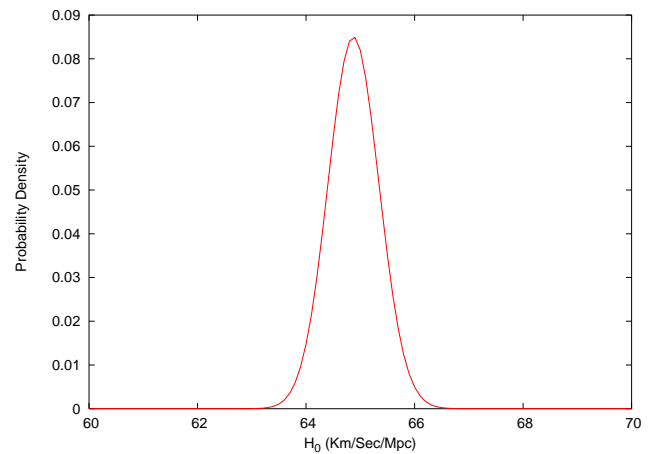}
 \caption{Probability density of the best reconstructed result from
 Gold data for different values of Hubble parameter. }
\label{h_margin_gold}
\end{figure}

By marginalizing over the Hubble parameter, we carry out a similar
treatment on the data as it has been done by \citet{riess06}
to calculate the $\chi^2$ for different cosmological models. As 
the Gold data are based on a Hubble parameter of 65 km/sec/Mpc,
the reconstruction method should be able to recover this value
for the Hubble constant. In fact the peak of the probability density of the
reconstructed result for different values of the
Hubble parameter should be close to $H_0=$ 65 km/sec/Mpc. In Figure
\ref{h_margin_gold} we show the probability density of the
best reconstructed result from Gold data for different values of the Hubble
parameter. We see that the probability density has a sharp peak around $H_0=$65
km/sec/Mpc.

We should also note here, that the reduced $\chi^2$ of the
reconstructed results seems to be consistently below $1$ (however it
is  not trivial to define the degree of freedom in our smoothing
method and hence the reduced $\chi^2$, but we can see that the
resultant $\chi^2$ of the reconstructed results is around $25$ less
than the number of data points). We can also see in Figure
\ref{chi_gold} that the reduced $\chi^2$ of the first initial
guess model, which is a $\Lambda$CDM model, is also below $1$. It
shows that the error-bars of the supernovae data points are quite
large and many different reconstructed results may have a reduced
$\chi^2$ of less than $1$. In this paper we only calculate the
$\chi^2$ of the reconstructed results and we compare different results by
calculating the $\Delta \chi^2$ to the best result with a 
minimum $\chi^2$.

\begin{figure*} 
\centering
\begin{center} 
\vspace{-0.05in}
\centerline{\mbox{\hspace{0.in} \hspace{2.1in}  \hspace{2.1in} }}
$\begin{array}{@{\hspace{-0.4in}}c@{\hspace{0.3in}}c@{\hspace{0.3in}}c}
\multicolumn{1}{l}{\mbox{}} &
\multicolumn{1}{l}{\mbox{}} &
\multicolumn{1}{l}{\mbox{}} \\ [-0.5cm]
 
\includegraphics[scale=0.38, angle=-90]{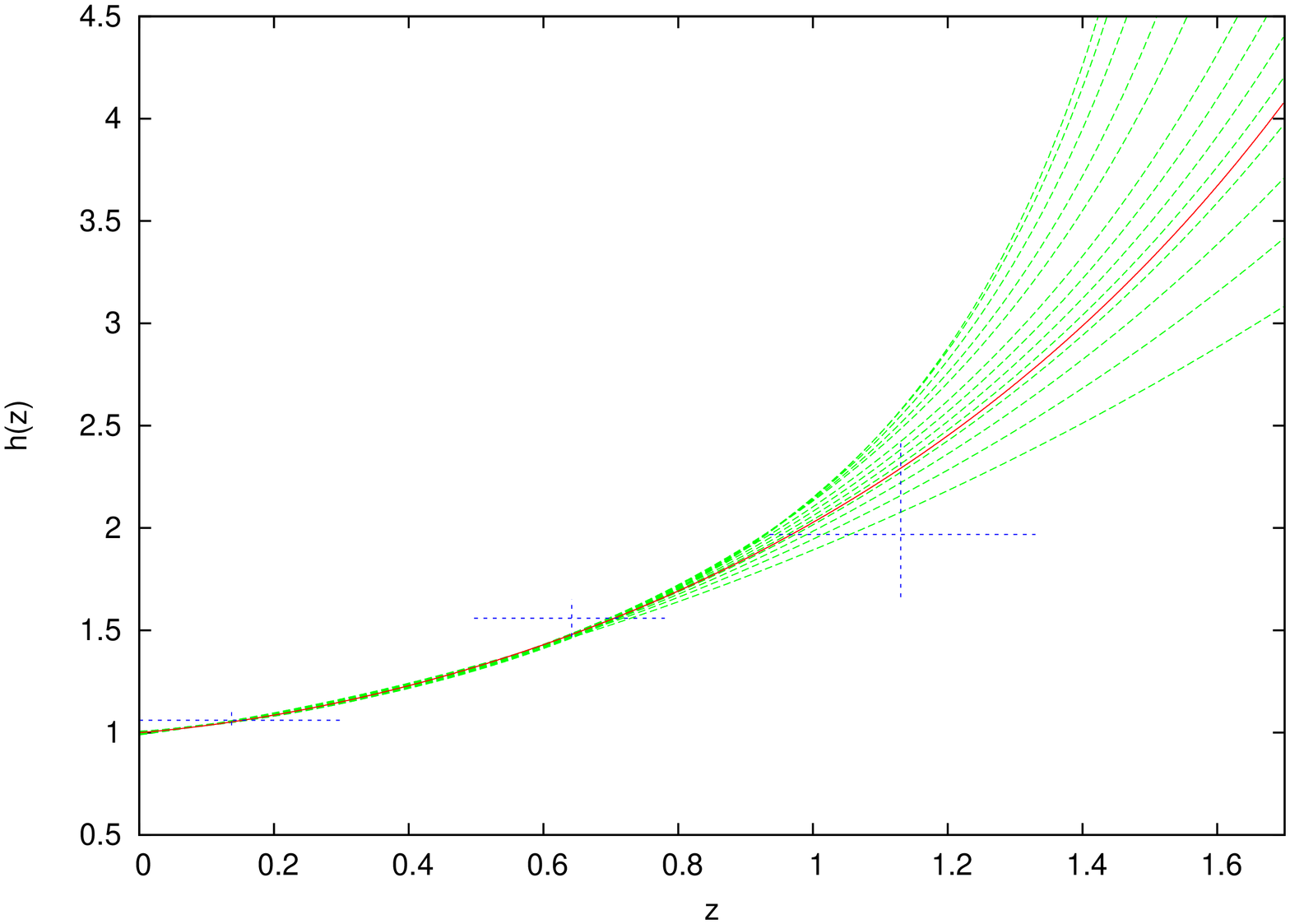}
 
\includegraphics[scale=0.38, angle=-90]{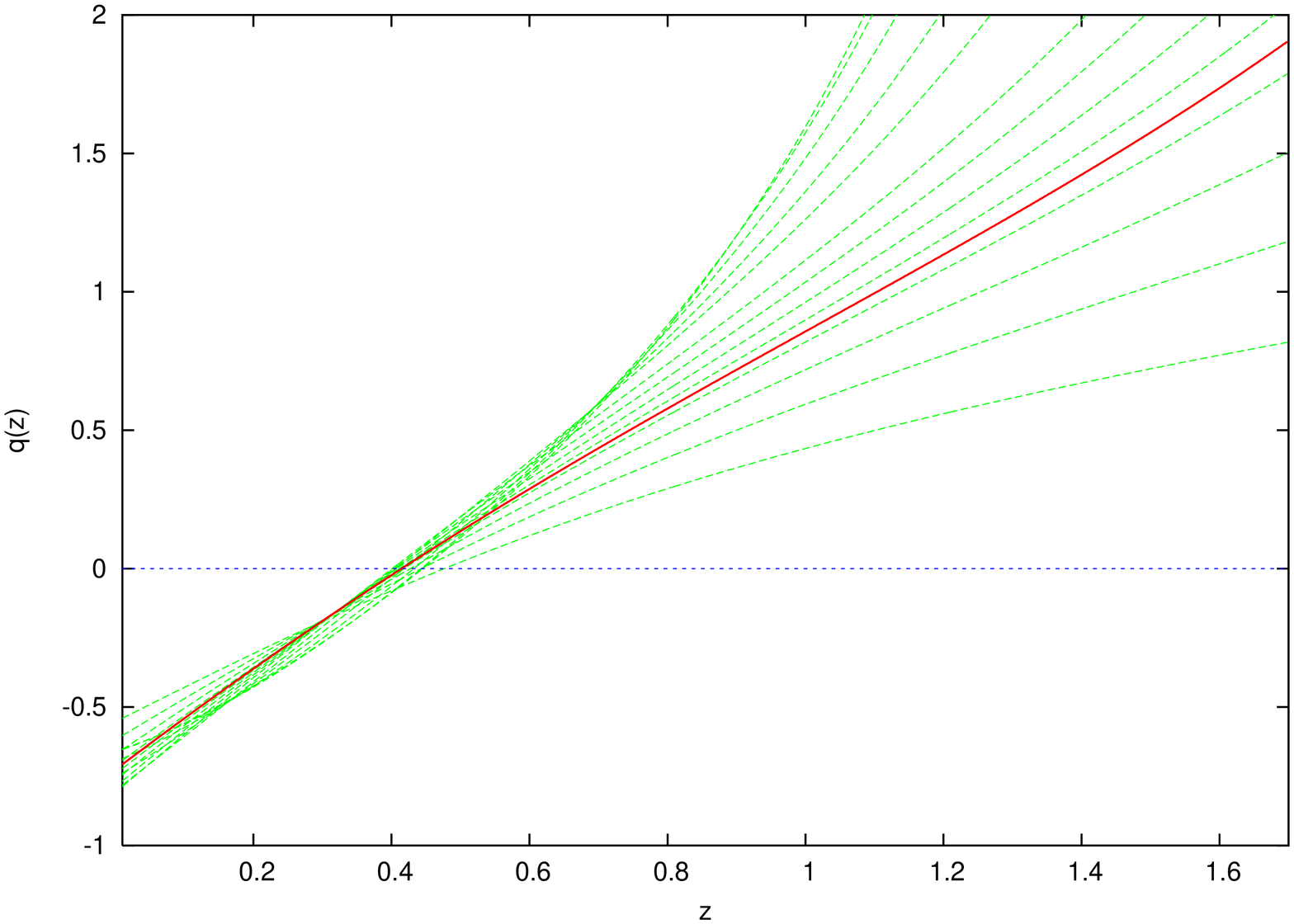}
\end{array}$

\end{center}
\caption{\small {Reconstructed $h(z)$ (left) and $q(z)$ (right) by
using Gold dataset. Red solid line is the best recovered result and
the green dashed lines are within $1\sigma$ away from the best
result. Based on our results, the transition between deceleration and
acceleration phases of the universe occurs at $0.38<z<0.48$ within
$1\sigma$ error-bar from the best recovered result. In the left panel
we can also see 3 uncorrelated and independent measurements of $h(z)$
from the Gold sample (blue dotted crosses from \citet{riess06}) 
for comparision with our reconstructed results.}}
\label{h_gold}
\end{figure*}
\begin{figure*} 
\centering
\begin{center} 
\vspace{-0.05in}
\centerline{\mbox{\hspace{0.in} \hspace{2.1in}  \hspace{2.1in} }}
$\begin{array}{@{\hspace{-0.4in}}c@{\hspace{0.3in}}c@{\hspace{0.3in}}c}
\multicolumn{1}{l}{\mbox{}} &
\multicolumn{1}{l}{\mbox{}} &
\multicolumn{1}{l}{\mbox{}} \\ [-0.5cm]
 
\includegraphics[scale=0.38, angle=-90]{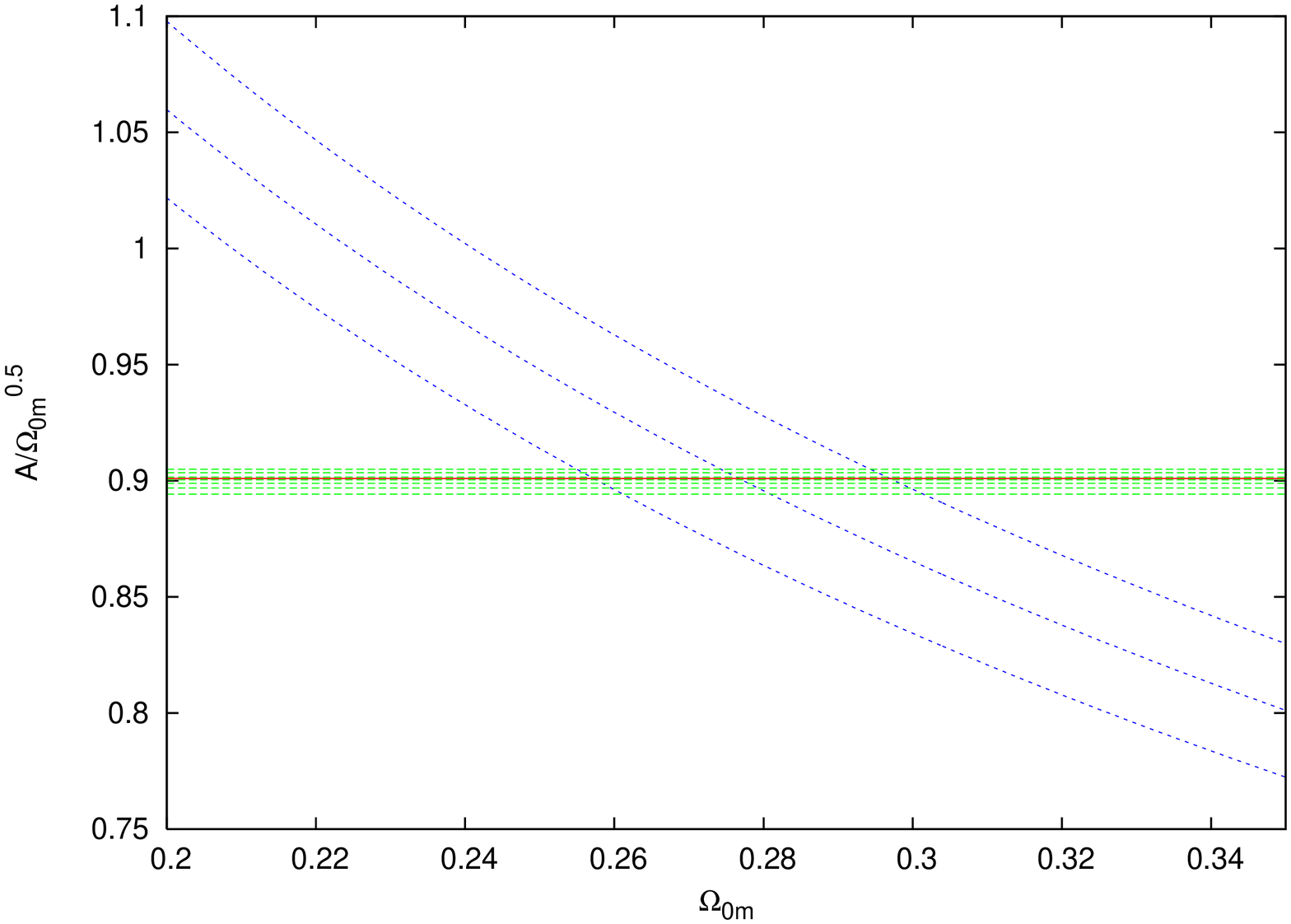}
 
\includegraphics[scale=0.38, angle=-90]{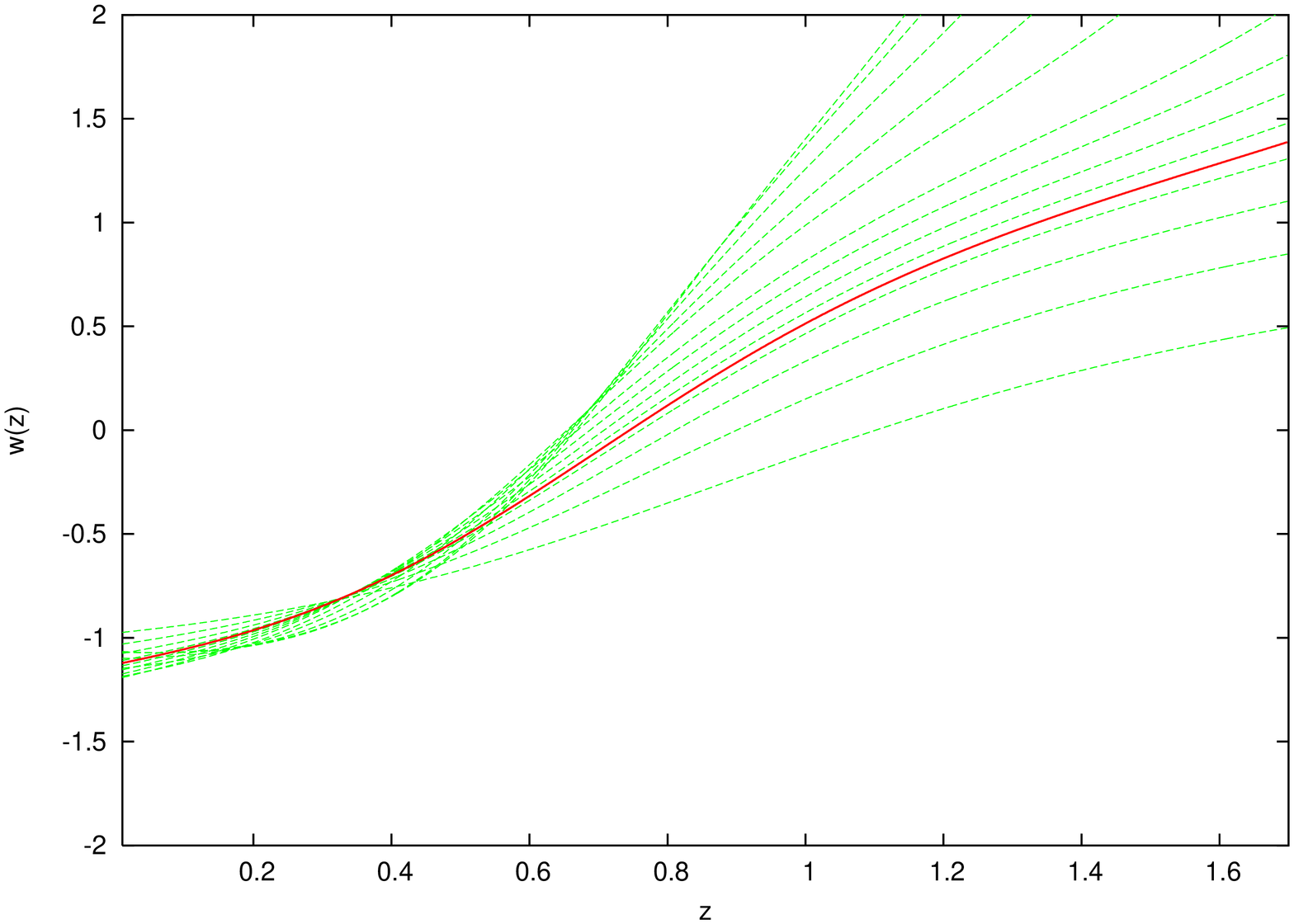}
\end{array}$

\end{center}
\caption{ Left panel: The derived value of $A/ \sqrt{\Omega_{0m}}$ from supernovae Gold data within its $1\sigma$ error-bars (red solid line and green dashed lines) in comparison with its measured value from observation of LRGs within its $1\sigma$ error-bars (blue dotted lines) for different values of $\Omega_{0m}$.
Right panel: Reconstructed $w(z)$ for the Gold dataset. Red solid line is the best recovered result and the green dashed lines are within $1\sigma$ away from the best result. To get these results, we have marginalised over $\Omega_{0m} = 0.277 \pm 0.022$.}
\label{w_gold}
\end{figure*}

In Figure \ref{h_gold}, left panel, we show the reconstructed $h(z)$
for the Gold data set. The red solid line has the highest likelihood
and is our best reconstruction. All the other lines are within
$1\sigma$ away from the best recovered result. These lines are
recovered results from our smoothing method by using different numbers
of iterations in the boot-strapping process. The $\Delta \chi^2$ for
all of these lines is less than $1$, and so we can consider them to
lie within $1\sigma$ of the best result. We should note that these
green dashed lines in Figure \ref{h_gold} are in fact a non-exhaustive
sample of results which are within $1\sigma$ away from the best
recovered result. As we see in the Figure \ref{h_gold}, the
reconstructed $h(z)$ at high redshift has a very big degeneracy. This
is expected since there is only a single supernova beyond redshift
$1.4$!

In this figure we can also see 3 uncorrelated and independent
measurements of $h(z)$ from the Gold sample (blue dotted crosses from
\citet{riess06}) for comparison with our results. We can see that
these two results are consistent with each other within their $1\sigma$ limits.
However we should mention here that in the \citet{wanteg05} method used by
\citet{riess06} for uncorrelated estimates of the expansion history, there is a
slight bias in the reconstruction of $h(z)$ and the higher derivatives
of the data. It is mainly because of using the average of the measured
quantity $h(z)^{-1}$, which typically is not a straight line. So as it
has been mentioned in \citet{wanteg05}, the  measured average of
$h(z)^{-1}$ (and hence $h(z)$) over a redshift bin  will generally lie
either slightly above or below the actual curve at the bin
center. This can be the reason that why the centres of the crosses in
Figure \ref{h_gold}, left panel, for uncorrelated estimates of the
expansion history, are slightly above or below our reconstructed
curve for the $h(z)$.

To reconstruct the Hubble parameter, $h(z)$, we do not need to know the value of $\Omega_{0m}$. Another important cosmological quantity which we can derive from the reconstructed $h(z)$ (independent of the value of $\Omega_{0m}$), is the deceleration parameter, $q(z)$,

\be
\label{eq:q}
q(z)=(1+z)\frac{H^{\prime}(z)}{H(z)} -1.
\ee

 In Figure \ref{h_gold}, right panel, we show reconstructed $q(z)$. For the Gold data our method shows that the transition between deceleration and acceleration occurred at $0.38 < z < 0.48$ (at $1\sigma$). The best reconstruction shows the redshift of transition to be $z_a \simeq 0.42$. This is in agreement with results obtained using parametric methods \citep{ass07,gw07}.

To derive the equation of state of dark energy $w(z)$, one needs to know the value of $\Omega_{0m}$, as we see in Eq.\ref{eq:w}. To estimate the value of $\Omega_{0m}$, without using any parameterisation and in a model-independent way, we can use the results of the detection of the baryon acoustic oscillation peak \citep{linderbao,bao05}. The distance factor $A$ up to redshift $0.35$, measured by observation of luminous red galaxies in detection of baryon acoustic oscillation peak (which have been claimed to be relatively independent of the model of dark energy), can be derived directly for different values of $\Omega_{0m}$ by using the reconstructed $h(z)$,
\be
A = \frac{\sqrt{\Omega_{0m}}}{h(z_1)^{1/3}}~\bigg\lbrack ~\frac{1}{z_1}~\int_0^{z_1}\frac{dz}{h(z)}
~\bigg\rbrack^{2/3}~,
\ee
where the measured value of $A$ is $A=0.469(\frac{n}{0.98})^{-0.35} \pm 0.017$ at $z_1=0.35$. The 3-year WMAP results, when combined with the results of baryon acoustic oscillations, yield $n=0.951$ for the spectral index of the primordial power spectrum \citep{spergel06,lambda}. By using the best reconstructed results for $h(z)$, we get $A/ \sqrt{\Omega_m} = 0.901$. In Figure \ref{w_gold}, left panel, we see the derived value of $A/ \sqrt{\Omega_m}$ from supernovae data in comparison with its measured value from observation of LRGs for different values of $\Omega_{0m}$. It is clear that these two independent observations which are completely different by nature, are very much in agreement if $0.255 < \Omega_{0m} < 0.299$. This derived value of $\Omega_{0m}$ is completely independent of any dark energy model assumption (within the framework of standard general relativity) and is in very close agreement with the results from large scale structure measurements from 2dF \citep{teg04} and SDSS \citep{col05}. This derived value of $\Omega_{0m}$ is also in good agreement with the results from \citet{reza06}, where a different model-independent method of reconstruction has been used.     
\begin{figure}
\includegraphics[scale=0.34, angle=-90]{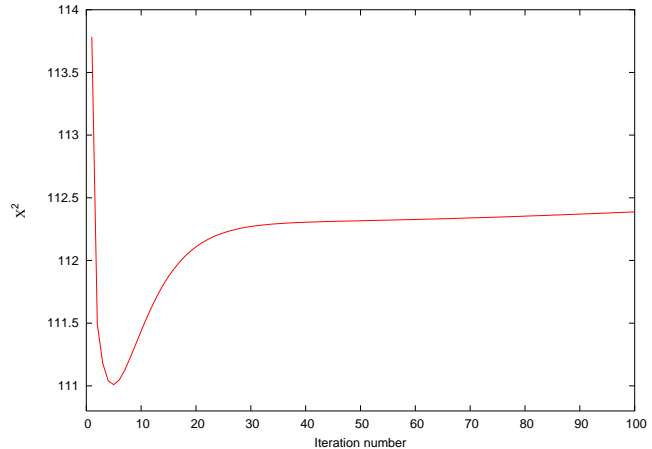}
 \caption{ Computed $\chi^2$ for the reconstructed results at each iteration for the SNLS dataset.}
\label{chi_snls}
\end{figure}

Now by marginalising over $\Omega_{0m}=0.277 \pm 0.022$, which is the range of agreement between the two observations, we can reconstruct $w(z)$ from our previously reconstructed $h(z)$. In Figure \ref{w_gold}, right panel, we show the reconstructed $w(z)$, marginalised over $\Omega_{0m}$ for the Gold dataset. We see that the data prefer evolving dark energy to the cosmological constant. The degeneracy for the equation of state of dark energy at high redshifts is very large and it is almost impossible to say much about $w(z)$ at high redshifts.

\section{Results from the SNLS dataset}

In this section we use the same procedure as we used in the previous section to deal with SNLS supernovae data. The SNLS dataset contains 115 data points in the range of $0.1<z<1.0$. We use this dataset, first to reconstruct the Hubble parameter, $h(z)$, and the deceleration factor, $q(z)$, up to redshift $1$. Then by using the results of detection of baryon acoustic oscillation peak we derive the value of $\Omega_{0m}$, following which we recover the form of $w(z)$. We use the distance modules of the supernovae available in Tables 8 and 9 in \citet{snls05} as our dataset in this section.
 
\begin{figure*} 
\centering
\begin{center} 
\vspace{-0.05in}
\centerline{\mbox{\hspace{0.in} \hspace{2.1in}  \hspace{2.1in} }}
$\begin{array}{@{\hspace{-0.4in}}c@{\hspace{0.3in}}c@{\hspace{0.3in}}c}
\multicolumn{1}{l}{\mbox{}} &
\multicolumn{1}{l}{\mbox{}} &
\multicolumn{1}{l}{\mbox{}} \\ [-0.5cm]
 
\includegraphics[scale=0.38, angle=-90]{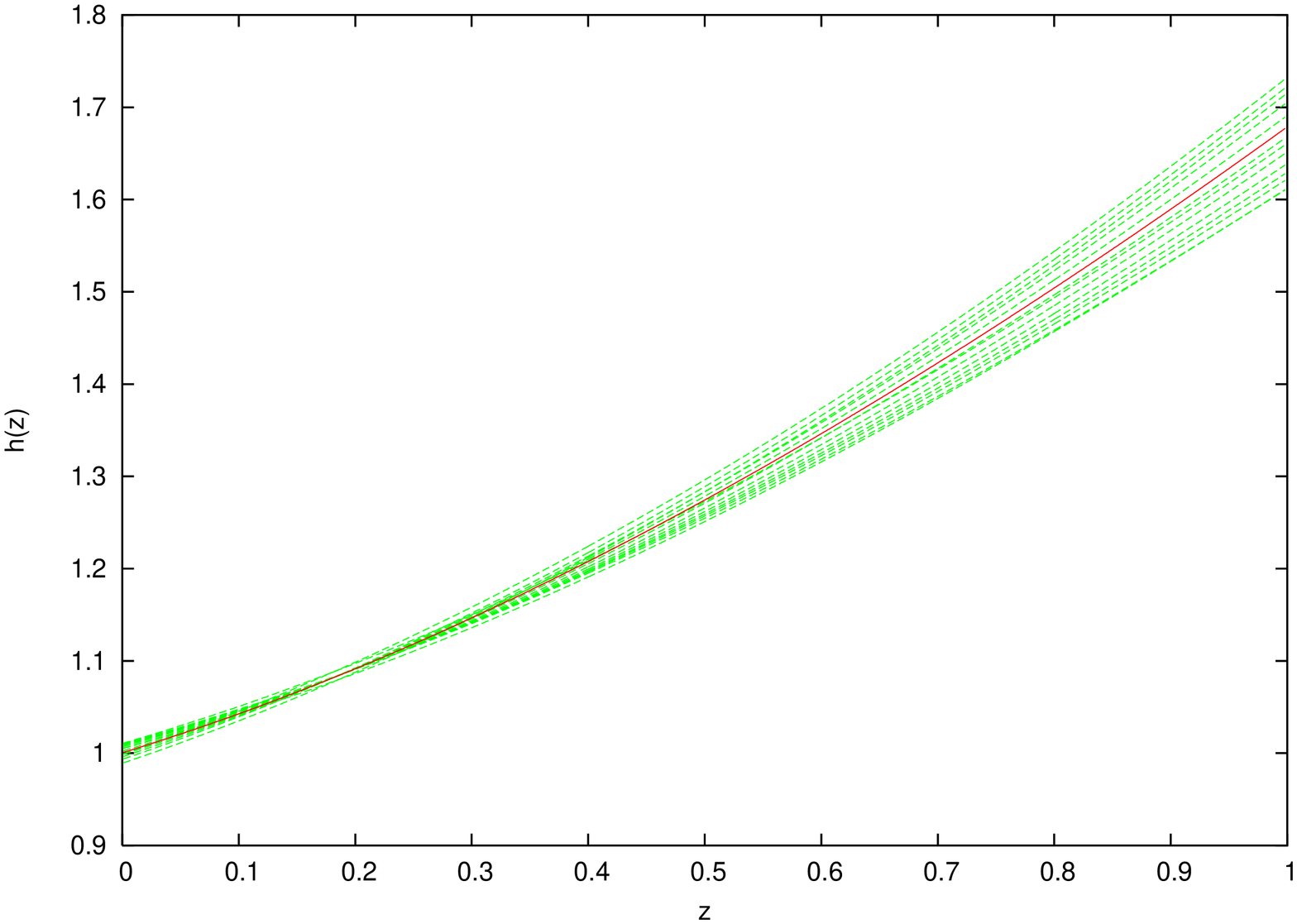}
 
\includegraphics[scale=0.38, angle=-90]{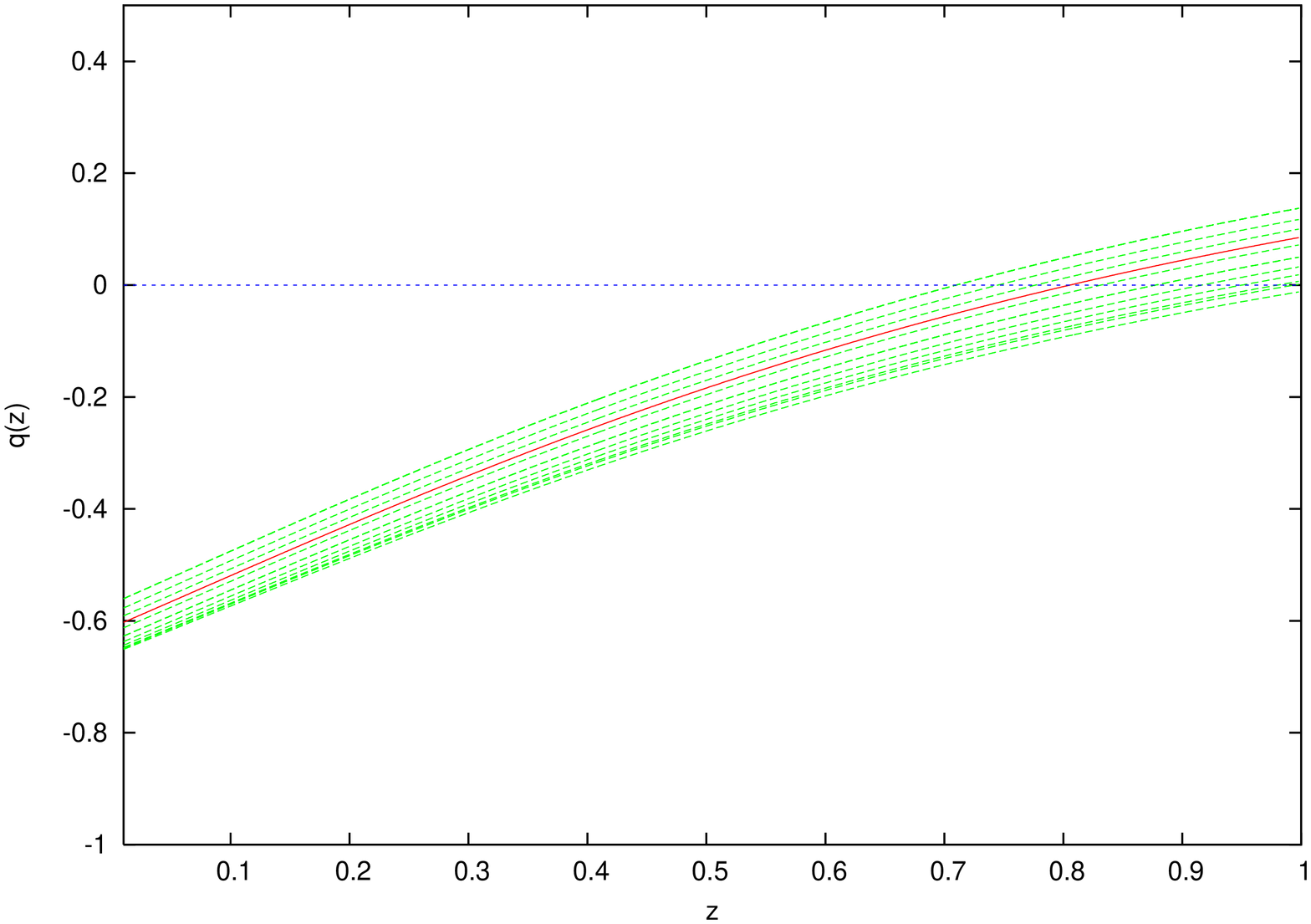}
\end{array}$

\end{center}
\caption{\small {Reconstructed $h(z)$ (left) and $q(z)$ (right) by using SNLS dataset. Red solid line is the best recovered result and the green dashed lines are within $1\sigma$ away from the best result. Based on our results, the transition between deceleration and acceleration phases of the universe occurs at $z>0.70$ within $1\sigma$ error-bar from the best recovered result.}}
\label{h_snls}
\end{figure*}

In Figure \ref{chi_snls} we see the computed  $\chi^2$ for the reconstructed results using smoothing method at each iteration. As we see, the $\chi^2$ diverges to its minimum value very fast at just the 5th iteration. In Figure \ref{h_snls} we show the reconstructed $h(z)$ (left panel) and $q(z)$ (right panel) for the SNLS dataset. The red solid line has the best likelihood, which is our best reconstructed result. All the other lines are within $1\sigma$ away from the best recovered result. We should like to emphasise here that these results (green dashed lines) are not representative of all the possibilities which give the likelihood within $1\sigma$ of the best recovered result. However they can show the overall behaviour of the quantities which we have studied. Our results for SNLS data show that the transition from deceleration to acceleration phase of the universe occurs at redshifts higher than $0.7$. The fact that we cannot put an upper limit to the redshift of the commence of acceleration is due to the absence of supernovae data at $z>1$ in SNLS dataset.

\begin{figure*} 
\centering
\begin{center} 
\vspace{-0.05in}
\centerline{\mbox{\hspace{0.in} \hspace{2.1in}  \hspace{2.1in} }}
$\begin{array}{@{\hspace{-0.4in}}c@{\hspace{0.3in}}c@{\hspace{0.3in}}c}
\multicolumn{1}{l}{\mbox{}} &
\multicolumn{1}{l}{\mbox{}} &
\multicolumn{1}{l}{\mbox{}} \\ [-0.5cm]
 
\includegraphics[scale=0.38, angle=-90]{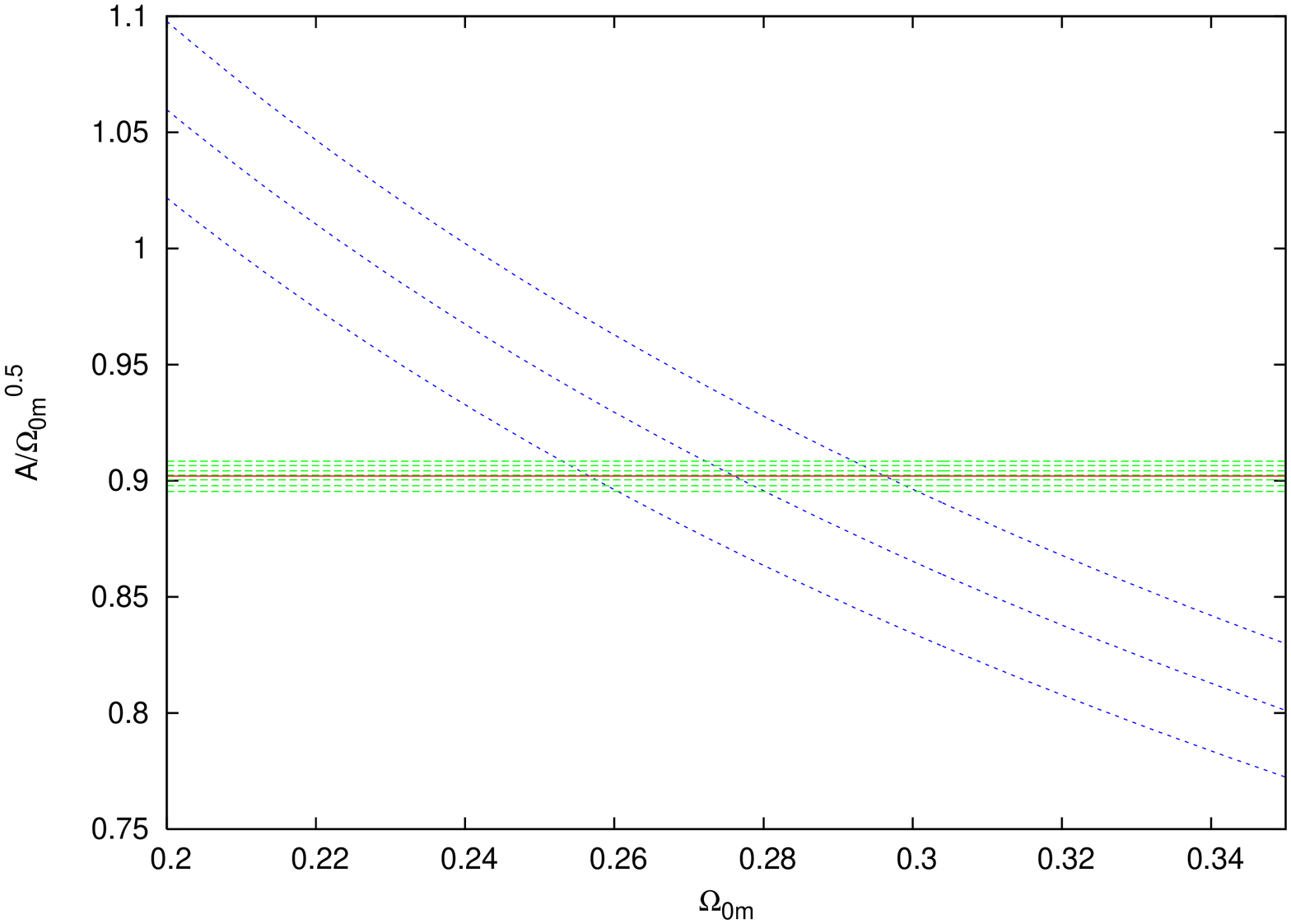}
 
\includegraphics[scale=0.38, angle=-90]{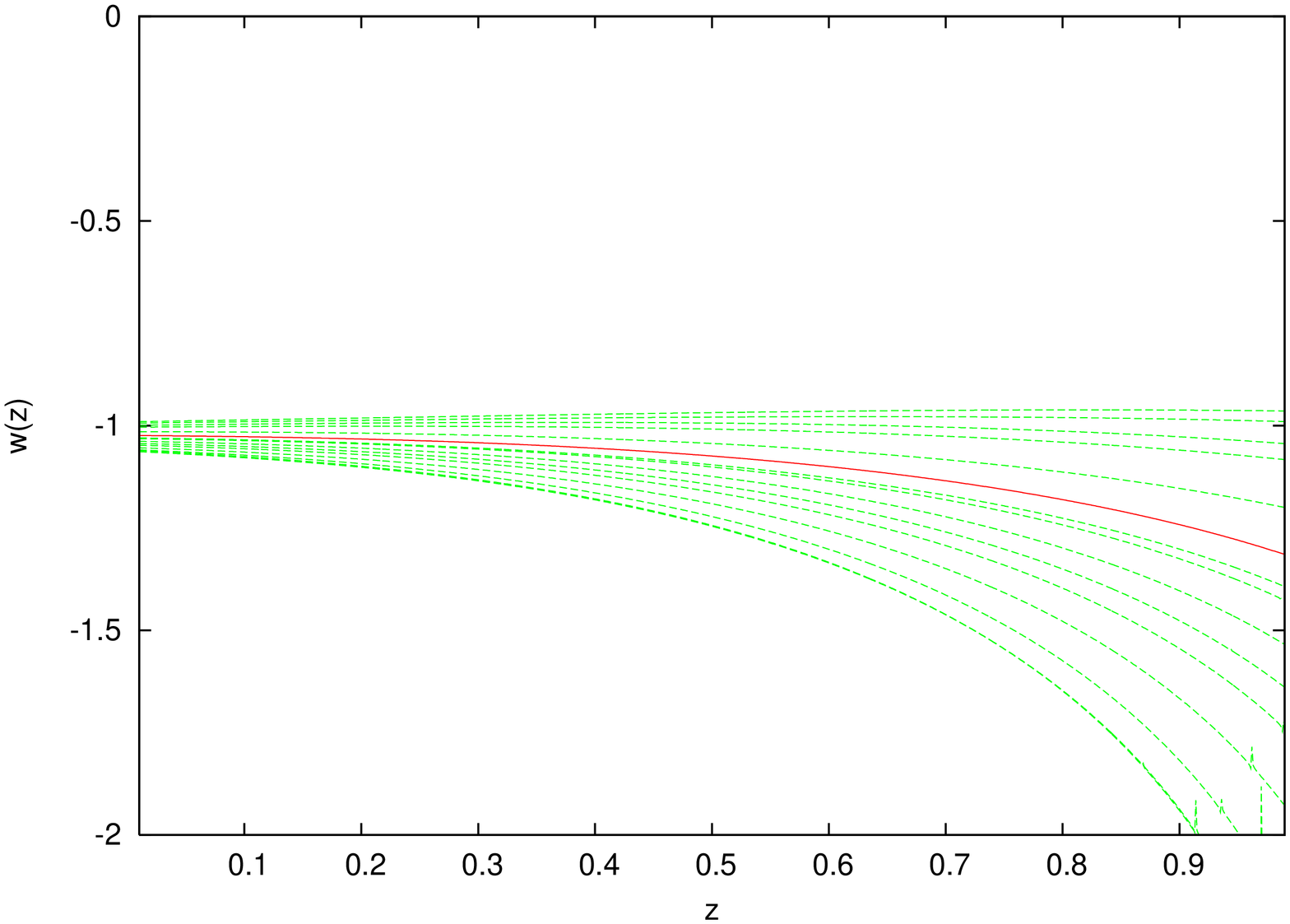}
\end{array}$

\end{center}
\caption{ Left panel: The derived value of $A/ \sqrt{\Omega_{0m}}$ from supernovae SNLS data within its $1\sigma$ error-bars (red solid line and green dashed lines) in comparison with its measured value from observation of LRGs within its $1\sigma$ error-bars (blue dotted lines) for different values of $\Omega_{0m}$. The consistency of this result with the result from the Gold sample is obvious.
Right panel: Reconstructed $w(z)$ for the SNLS dataset. Red solid line is the best recovered result and the green dashed lines are within $1\sigma$ away from the best result. To get these results, we have marginalised over $\Omega_{0m} = 0.276 \pm 0.023$.}
\label{w_snls}
\end{figure*}

As we have discussed earlier in the previous section, we use the
results of detection of baryon acoustic oscillation peak to determine
the value of $\Omega_{0m}$. Then by marginalising over the recovered
value of $\Omega_{0m}$, we derive the dynamics of $w(z)$. In Figure
\ref{w_snls} we see the derived value of $\Omega_{0m}$ and the
reconstructed form of $w(z)$. We see that the $\Lambda$CDM model is in
much better agreement with SNLS data than with Gold data.

By comparing the recovered results from the SNLS and Gold datasets,
we can clearly see an inconsistency between these two supernovae
datasets. This inconsistency is obvious by looking at the
reconstructed $q(z)$ and $w(z)$ in the middle
and high redshift ranges. Gold data suggest the redshift of the
commencement of acceleration at $z_a \simeq 0.42$ while SNLS data suggest
 $z_a \simeq 0.80$. The reconstructed $w(z)$ from these two datasets
also shows a very different behavior in the middle and high
redshift ranges. The discord between Gold and SNLS supernovae datasets has been
reported and studied earlier by \cite{nesper}, and a similar discord
between Gold supernovae data and other
cosmological observations like Cosmic Microwave Background
observations from WMAP and observations of cluster abundance, has also been
reported earlier by \cite{harvindar}. However SNLS supernovae data
seem to be in good agreement with the other cosmological
observations. Based on all these results and analyses, we may conclude
that some significant systematics in the Gold data (or in a part of
the data) might be the reasos for these inconsistencies.

Interestingly, the recovered values of $\Omega_{0m}$ from Gold
and SNLS data (by using the results of detection of baryon acoustic oscillation
peak), are in very close agreement. In both cases the
derived value of $\Omega_{0m}$ is around $0.276 \pm 0.022$.
We should note here that the two data sets rely on
pretty much the same nearby supernovae samples and that is why the results are
similar in this range. It is something which we logically expect to
get. But in fact it shows one of the advantages of this
method over the functional fitting methods. By using a functional
fitting method, the recovered results
at any redshift would be equally dependent on the data set in the
whole redshift range. But here, by using our smoothing method, we can
clearly see that despite the significant differences between the
reconstructed results from Gold and SNLS datasets in the middle and
high redshift regions, the reconstructed results for the expansion
history at low redshifts (which we use to estimate the value of matter
density), are not affected by the big differences between the two datasets
at the higher redshifts.


\section{Discussion and conclusion}

In this paper we have shown that by improving the efficacy of the smoothing method \citep{sass06}, we can reconstruct the expansion history of the universe in a model-independent way, using current supernovae data. 
We have used the smoothing method to reconstruct the expansion history
of the universe, $h(z)$, the deceleration parameter, $q(z)$, the value
of $\Omega_{0m}$ and the equation of state of dark energy, $w(z)$,
independently of any assumption of the theoretical model of the
universe, within the framework of standard general relativity. This is
an advantage of this method over the functional fitting methods where
the results are usually biased by the form of the functional fitting or the
assumed theoretical model. We dealt with two recent
datasets, Gold and SNLS in our analysis. In determining the value of
$\Omega_{0m}$, we found excellent agreement between Gold and SNLS
datasets. This determination is directly related to the supernovae
data points at redshifts lower than $z=0.35$. We have got $\Omega_{0m}
\approx 0.276 \pm 0.023$ for both Gold and SNLS datasets, which is in
good agreement with results of SDSS and 2dF large scale structure
observations, and also with results of recent Chandra X-ray
observations of the relaxed galaxy clusters \citep{allen}.

 This derived value of $\Omega_{0m}$ also agree with the recent WMAP 3
years CMB data, if we assume the broken scale invariant spectrum for
the form of the primordial spectrum \citep{prd07}. In the derivation
of $q(z)$ and the stage of transition from deceleration to
acceleration in the dynamics of the universe, we found disagreement
between Gold and SNLS datasets. Gold data suggest the redshift of the
commence of acceleration at $z_a \simeq 0.42$ while SNLS data suggest
$z_a \simeq 0.80$. 


 After marginalizing over the derived value of $\Omega_{0m}$, we have
reconstructed $w(z)$. The inconsistency between Gold and SNLS
supernovae datasets is also obvious by looking at the reconstructed $w(z)$ from
these two datasets. The derived form of $w(z)$ from SNLS dataset, is in good
concordance with $\Lambda$CDM model, while Gold dataset
prefers an evolving form of dark energy (however $\Lambda$CDM is
still in agreement with the Gold dataset to within $2\sigma$). This discrepancy
between Gold and SNLS datasets has been reported earlier by other groups
(\cite{nesper,ass07}). As the Gold sample is also relatively in disagreement
with the other cosmological observations like CMB and observations of cluster
abundance (\cite{harvindar}), we may conclude that the effect of
systematics in the Gold dataset (or at least in a part of the data) is
significant.

The large error-bars at the high redshifts for the reconstructed
results, reflect the significant lack of data points. This effect
may not be seen if we use some of the parametric methods of analysis,
but as we deal with the data directly here, we notice that the lack of
data points at high redshifts limits our ability to say much about the
behavior of the Universe at the early stages at high redshifts. This is another
important feature of our smoothing method in which the reconstructed
results at any redshift rely mostly on the supernovae data points at
the same redshift range.

\section{Acknowledgement}     

I would like to thank Varun Sahni, Alexei Starobinsky, Tarun Souradeep, Eric Linder, Tarun Deep Saini, Stephane Fay, Arnab Kumar Ray and Ujjaini Alam for useful discussions.

\appendix
\section{examining the robustness of the method}

In this section we show that the results of the smoothing method are robust against the choice of the initial guess model and also to the chosen value of $\Delta$. 

\begin{figure*} 
\centering
\begin{center} 
\vspace{-0.05in}
\centerline{\mbox{\hspace{0.in} \hspace{2.1in}  \hspace{2.1in} }}
$\begin{array}{@{\hspace{-0.4in}}c@{\hspace{0.3in}}c@{\hspace{0.3in}}c}
\multicolumn{1}{l}{\mbox{}} &
\multicolumn{1}{l}{\mbox{}} &
\multicolumn{1}{l}{\mbox{}} \\ [-0.5cm]
 
\includegraphics[scale=0.38, angle=-90]{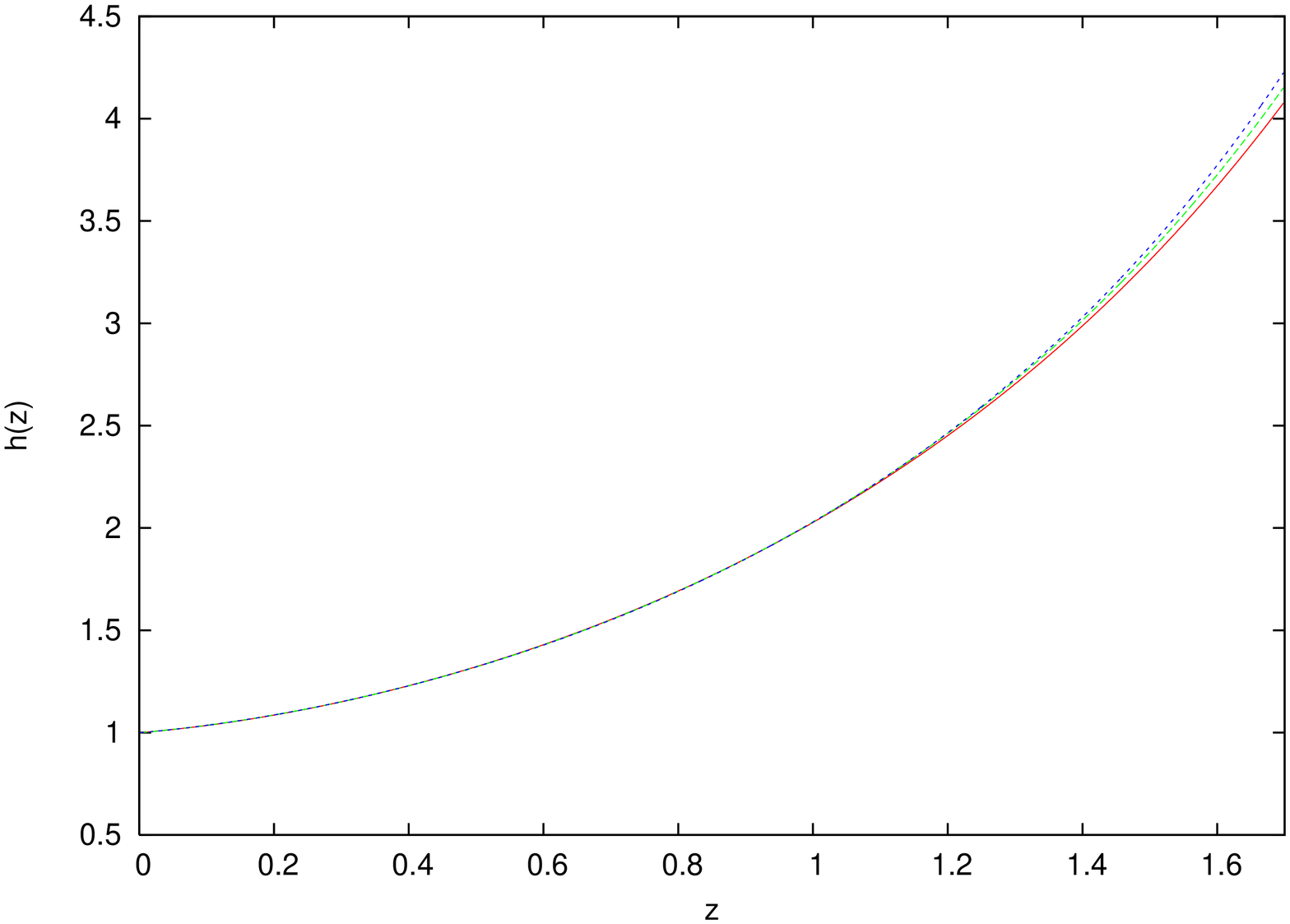}
 
\includegraphics[scale=0.38, angle=-90]{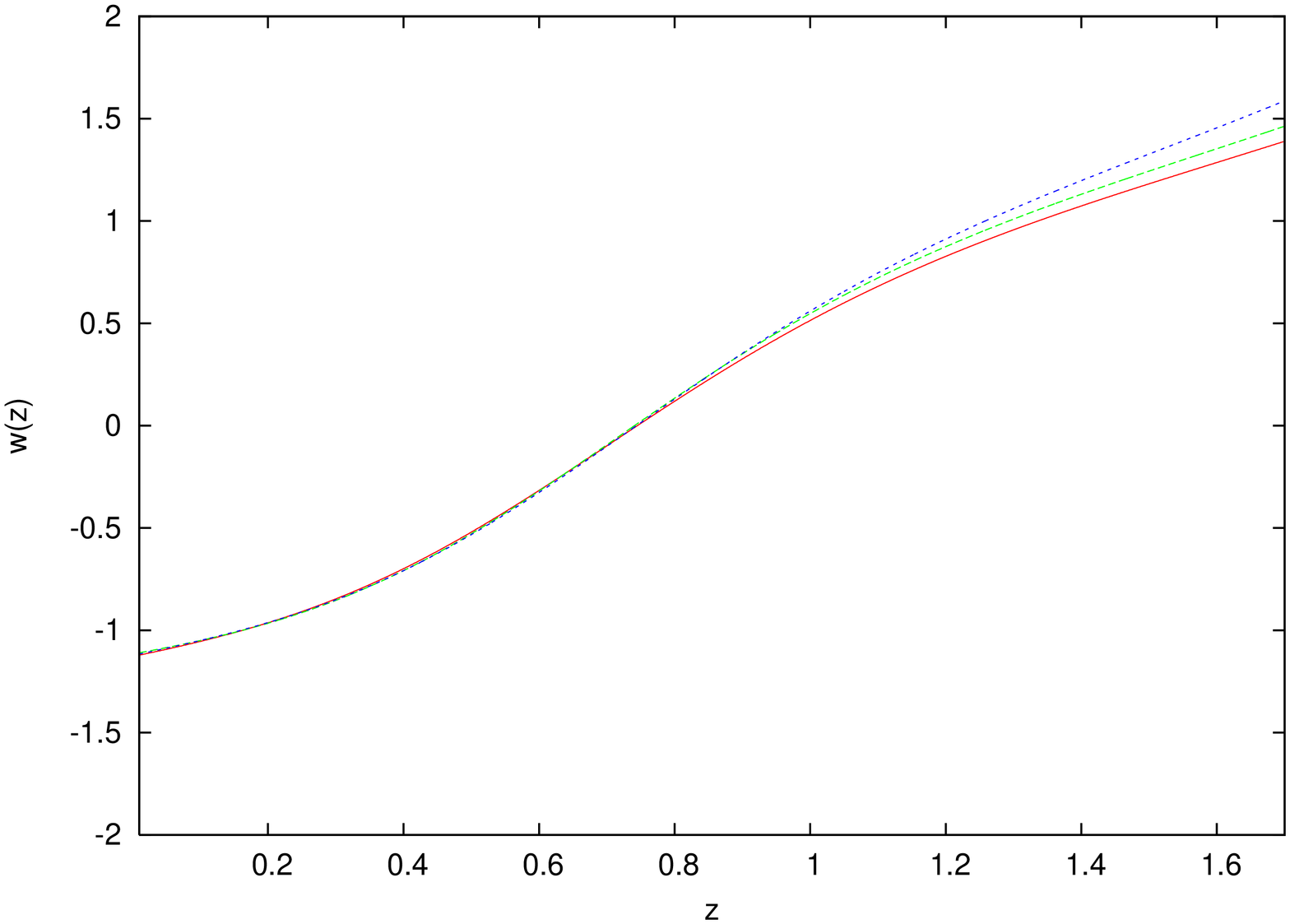}
\end{array}$

\end{center}
\caption{ Reconstructed $h(z)$ (left panel) and $w(z)$ (right panel) for the Gold dataset by assuming three different initial guess models. The red solid line is the reconstructed result by using a flat $\Lambda$CDM model with $\Omega_{0m}=0.30$ as the initial guess model. The green dashed line is the reconstructed results by using a flat $\Lambda$CDM model with $\Omega_{0m}=0.25$, and the blue dotted line is the reconstructed result by using a flat quiessence model with $w(z)=-0.8$ and $\Omega_{0m}=0.30$ as the initial guess models. We can clearly see that the results are almost identical which shows the robustness of the method for the different choices of the initial guess model.}
\label{GT_fig}
\end{figure*}

We assumed three different cosmological models as our initial guess model and we applied our smoothing method on the Gold dataset. The final results by using these three different initial guess models are almost identical with $\Delta \chi^2 < 0.01$. We have got $\chi^2 = 157.40$ by using a flat $\Lambda$CDM model with $\Omega_{0m} =0.30$ as the initial guess model after 89 iteration, while we have got $\chi^2 = 157.40$ for a flat $\Lambda$CDM model with $\Omega_{0m} =0.25$ after 91 iteration, and $\chi^2 = 157.39$ for a flat quiessence model with $\Omega_{0m} =0.30$ and $w(z)=-0.8$ after 104 iteration. In figure \ref{GT_fig} we can see the reconstructed $h(z)$ and $w(z)$ for the Gold data set by assuming these three different initial guess models. As we see, the robustness of the method for the choice of the initial guess model is obvious.

\begin{figure*} 
\centering
\begin{center} 
\vspace{-0.05in}
\centerline{\mbox{\hspace{0.in} \hspace{2.1in}  \hspace{2.1in} }}
$\begin{array}{@{\hspace{-0.4in}}c@{\hspace{0.3in}}c@{\hspace{0.3in}}c}
\multicolumn{1}{l}{\mbox{}} &
\multicolumn{1}{l}{\mbox{}} &
\multicolumn{1}{l}{\mbox{}} \\ [-0.5cm]
 
\includegraphics[scale=0.38, angle=-90]{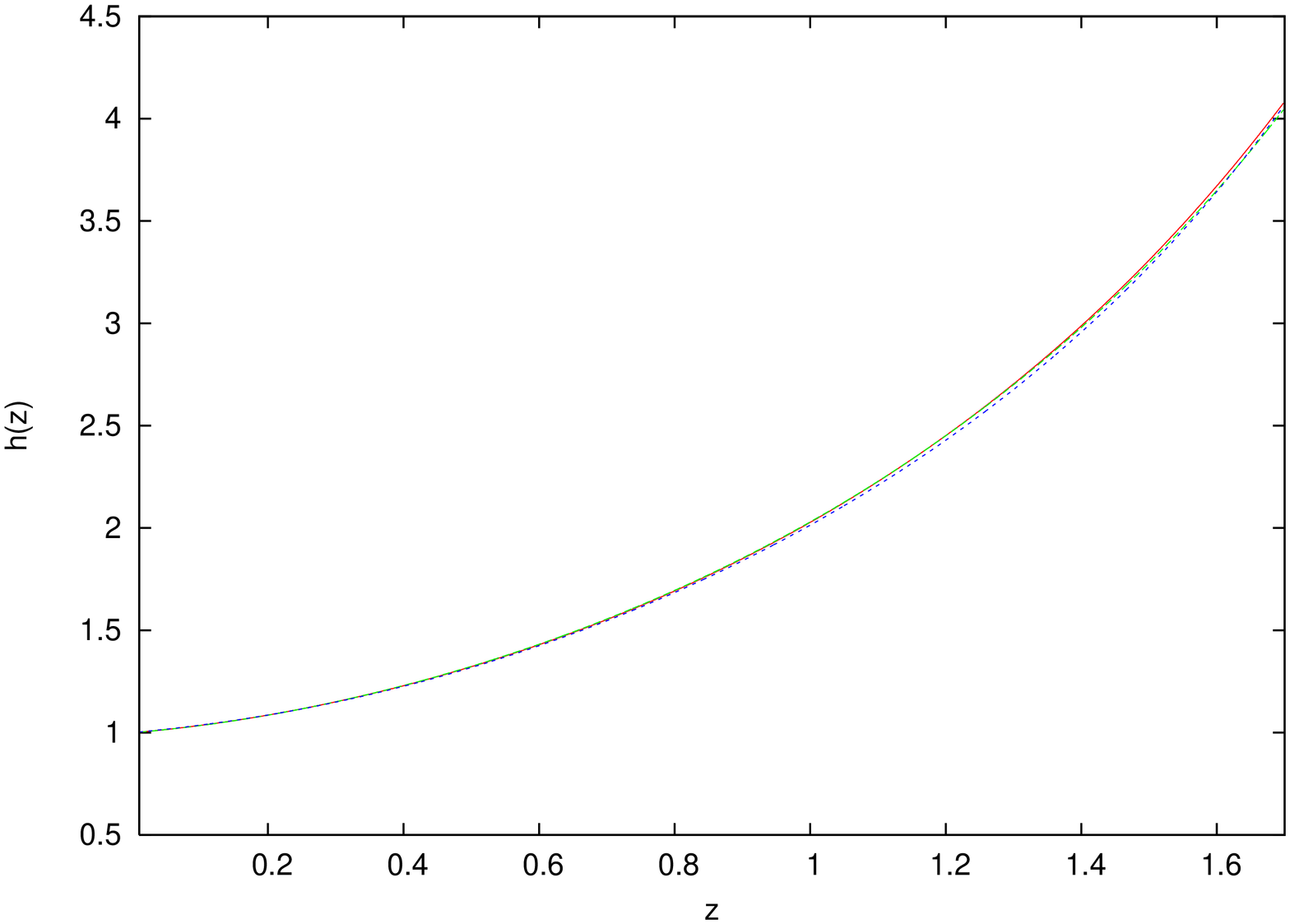}
 
\includegraphics[scale=0.38, angle=-90]{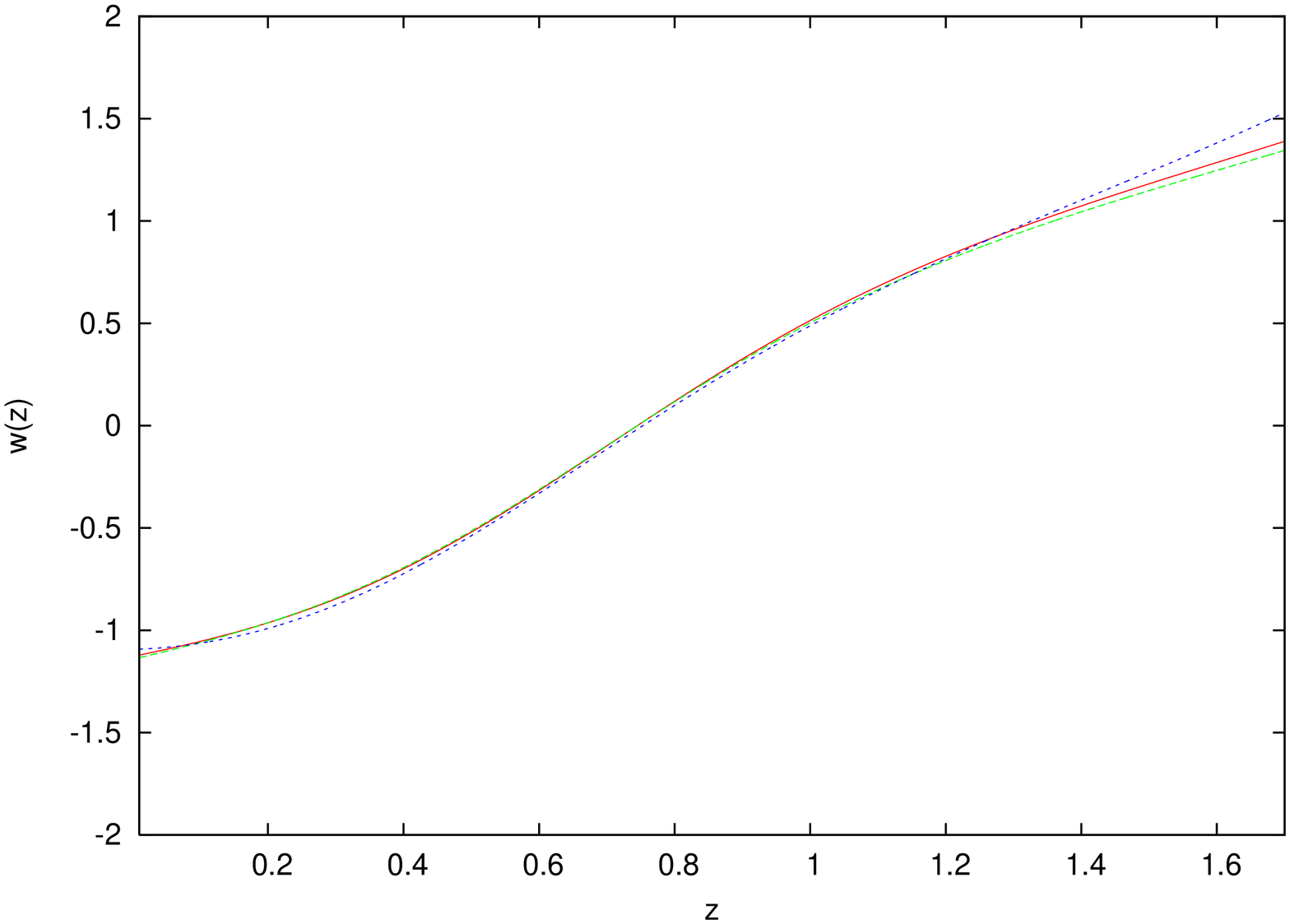}
\end{array}$

\end{center}
\caption{ Reconstructed $h(z)$ (left panel) and $w(z)$ (right panel) for the Gold dataset by using three different values of $\Delta$ (width of smoothing). The red solid line is the reconstructed result by using $\Delta = 0.60$. The green dashed line is the reconstructed results by using $\Delta = 0.90$, and the blue dotted line is the reconstructed result by using $\Delta = 0.30$. In all these cases we have stopped the boot-strapping process after reaching to the minimum $\chi^2$. We can see that the method is robust against the variation of $\Delta$ in a wide range.}
\label{WT_fig}
\end{figure*}

We have also used different values of $\Delta$ (width of smoothing in Eq.\ref{eq:bg}), in our reconstruction process to check the reliability and stability of our results against the changes in the value of $\Delta$. We have used three values of $\Delta$ equal to $0.30,0.60$ and $0.90$ in our smoothing method and we have applied it on the Gold dataset. By using $\Delta = 0.30$ we have got $\chi^2 = 157.38$ after 9 iteration, while we have got $\chi^2 = 157.40$ by using $\Delta = 0.60$ after 89 iteration, and $\chi^2 = 157.41$ by using $\Delta = 0.90$ after 407 iteration. In figure \ref{WT_fig} we can see the reconstructed $h(z)$ and $w(z)$ for the Gold dataset by using these three values of $\Delta$. We can clearly see that the results are not sensitive to the given value of $\Delta$. These two examinations confirm the overall robustness of the method for different initial assumptions.

\end{document}